\documentclass[lettersize,journal]{IEEEtran}
\usepackage{amsmath,amsfonts}
\usepackage{algorithmic}
\usepackage{algorithm}
\usepackage{array}
\usepackage[caption=false,font=normalsize,labelfont=sf,textfont=sf]{subfig}
\usepackage{textcomp}
\usepackage{stfloats}
\usepackage{url}
\usepackage{verbatim}
\usepackage{graphicx}
\usepackage{cite}
 \usepackage{multirow}
\usepackage{makecell}
\usepackage{booktabs}
\usepackage[inkscapelatex=false]{svg}
\usepackage{tikz,xcolor,hyperref}% Make Orcid icon
\hypersetup{hidelinks, colorlinks=true, allcolors=black, pdfstartview=Fit, breaklinks=true}
\definecolor{lime}{HTML}{A6CE39}
\DeclareRobustCommand{\orcidicon}{%
    \begin{tikzpicture}
    \draw[lime, fill=lime] (0,0) 
    circle [radius=0.16] 
    node[white] {{\fontfamily{qag}\selectfont \tiny ID}};    \draw[white, fill=white] (-0.0625,0.095) 
    circle [radius=0.007];    \end{tikzpicture}
    \hspace{-2mm}}
\foreach \x in {A, ..., Z}{%
    \expandafter\xdef\csname orcid\x\endcsname{\noexpand\href{https://orcid.org/\csname orcidauthor\x\endcsname}{\noexpand\orcidicon}}
    }
\begin{document}

 \title{Federated User Preference Modeling for Privacy-Preserving Cross-Domain Recommendation}

 %code: FedPCL_MDR_TMM/FedPCL_MDR_v4_no_disen_personalize/lad_protection/logs_new

\author{Li Wang\orcidA{}, Shoujin Wang\orcidB{}, Quangui Zhang\orcidE{}, Qiang Wu\orcidD{}, Senior Member, IEEE, and Min Xu$^*$\orcidC{}, IEEE, Member
%~\IEEEmembership{Staff,~IEEE,}
%         <-this % stops a space
\thanks{Li Wang, Qiang Wu, and Min Xu are with the School of Electrical and Data Engineering, University of Technology Sydney, Sydney 2000, Australia. Shoujin Wang is with the institute of data science, University of Technology Sydney, Sydney 2000, Australia.  Quangui Zhang is with the School of Artificial Intelligence, Chongqing University of Arts and Sciences, Chongqing 402160, China.} \thanks{$*$Corresponding author: Min Xu (e-mail: Min.Xu@uts.edu.au)}}% <-this % stops a space

\markboth{Journal of \LaTeX\ Class Files,~Vol.~14, No.~8, August~2021}%
{Shell \MakeLowercase{\textit{et al.}}: A Sample Article Using IEEEtran.cls for IEEE Journals}


\maketitle
\begin{abstract}

Cross-domain recommendation (CDR) aims to address the data-sparsity problem by transferring knowledge across domains. Existing CDR methods generally assume that the user-item interaction data is shareable between domains, which leads to privacy leakage. Recently, some privacy-preserving CDR (PPCDR) models have been proposed to solve this problem. However, they primarily transfer simple representations learned only from user-item interaction histories, overlooking other useful side information, leading to inaccurate user preferences. Additionally, they transfer differentially private user-item interaction matrices or embeddings across domains to protect privacy. However, these methods offer limited privacy protection, as attackers may exploit external information to infer the original data. 
To address these challenges, we propose a novel Federated User Preference Modeling (FUPM) framework. In FUPM, first, a novel comprehensive preference exploration module is proposed to learn users' comprehensive preferences from both interaction data and additional data including review texts and potentially positive items. Next, a private preference transfer module is designed to first learn differentially private local and global prototypes, and then privately transfer the global prototypes using a federated learning strategy. These prototypes are generalized representations of user groups, making it difficult for attackers to infer individual information.
% Then, a private preference transfer module is designed to first learn local and global prototypes respectively and then privately transfer global prototypes with federated learning strategy. To further enhance privacy protection, the LDP technique is applied to local prototypes before transfer.
Extensive experiments on four CDR tasks conducted on the Amazon and Douban datasets validate the superiority of FUPM over SOTA baselines. 
Code is available at https://github.com/Lili1013/FUPM. 

\end{abstract}


\section{Introduction}

\IEEEPARstart{C}{ross}-domain recommendation (CDR) aims to address the data-sparsity problem by transferring cross-domain knowledge \cite{nie2023crossdomain, li2020ddtcdr, zhu2020graphical,lu2023contrastive}. Most existing CDR methods \cite{zhao2024domainoriented,li2020ddtcdr,zhao2024crossdomain} assume that user-item interaction data, such as user-item ratings and review texts, are shareable across different domains. However, since such data contains sensitive user information and is typically managed by different companies, sharing it has become increasingly unacceptable due to business competition and privacy protection laws like the GDPR\footnote{https://gdpr-info.eu}. Therefore, it is crucial to design a privacy-preserving cross-domain recommendation (PPCDR) method that provides high-quality recommendations while safeguarding user privacy.

In this paper, we focus on a general PPCDR problem. We assume that there are multiple domains, each containing user-item ratings and review texts. The data within each domain is private and cannot be shared across domains. Therefore, the PPCDR problem is more challenging compared to the conventional CDR problem. We aim to learn comprehensive user preferences and privately transfer them across domains to alleviate the data-sparsity problem while protecting privacy. 
%while protecting user privacy.

% \begin{figure}[!t]
% 	\centering
% \includesvg[width=0.5\textwidth]{motivation1.svg}
% 	\caption{(a) Traditional CDR methods typically map and transfer user embeddings directly, without considering user privacy. (b) Existing PPCDR methods rely solely on user-item interaction histories to learn user embeddings and transfer differentially private embeddings across domains. (c) FUPM first utilizes review texts and potentially positive items to learn comprehensive user preferences and then privately transfers these preferences using differentially private prototypes within the FL framework.}
%   \label{motivation}
%   \vspace{-5mm}
% \end{figure}

\begin{figure}[!t]
	\centering
\includegraphics[width=0.5\textwidth]{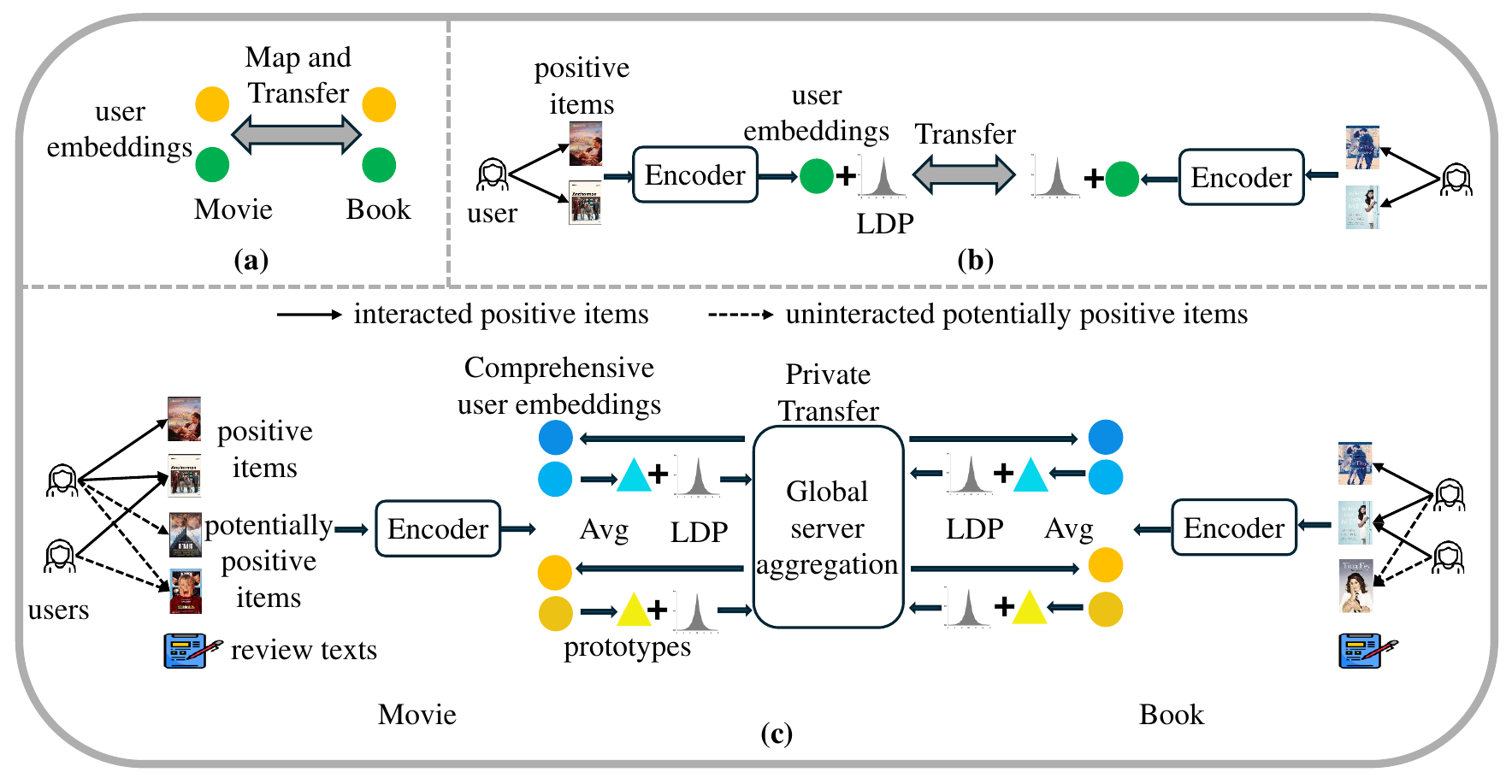}
	\caption{(a) Traditional CDR methods typically map and transfer user embeddings directly, without considering user privacy. (b) Existing PPCDR methods rely solely on user-item interaction histories to learn user embeddings and transfer differentially private embeddings across domains. (c) FUPM first utilizes review texts and potentially positive items to learn comprehensive user preferences and then privately transfers these preferences using differentially private prototypes within the FL framework.}
  \label{motivation}
  \vspace{-5mm}
\end{figure}

Existing studies do not effectively solve the PPCDR problem. Conventional CDR methods \cite{ijcai2019p587,zhu2020graphical,qian2018crossdomain,chen2022ecommerce} typically transfer original user-item interaction data or plaintext embeddings across domains, as shown in Fig. \ref{motivation} (a). For instance, DARec \cite{ijcai2019p587} transfers rating patterns between domains using deep learning techniques. GA-DTCDR \cite{zhu2020graphical} learns user and item embeddings in each domain and transfers user embeddings via an attention network. Despite achieving optimal recommendation performance, these methods may leak user privacy. To tackle this challenge, PPCDR methods \cite{chen2023winwin, chen2022differential, liao2023ppgencdr,liu2023federateda} have received significant attention in recent years. P2FCDR \cite{chen2023winwin} learns an orthogonal mapping matrix to transform embeddings and then utilizes the Local Differential Privacy (LDP) technique to protect user privacy. Similarly, PriCDR \cite{chen2022differential} uses LDP technology to publish the rating
matrix in the source domain and subsequently transfers the published matrix to the target domain. The authors in \cite{tian2024privacypreserving} propose a federated graph learning method that utilizes FL and LDP techniques to protect user privacy.  
However, these methods face two key challenges.

\noindent \textbf{CH1.} These methods only utilize user-item interaction histories to learn and transfer user and item embeddings, as illustrated in Fig. \ref{motivation} (b). They neglect informative review texts and \textit{potentially positive items} for exploring comprehensive user preferences. We refer to potentially positive items as items that a user has not interacted with but might be interested in. 

\noindent \textbf{CH2.} While these methods protect user privacy by sharing differentially private interaction matrices or embeddings, as shown in Fig. \ref{motivation} (b), they still exhibit limited privacy-preserving capacity. Attackers can leverage external information to infer sensitive data. For instance, if attackers have access to a public dataset, they can use it to learn reference user/item embeddings and then correlate these with the differentially private embeddings to infer individual user preferences.

To address these challenges, we propose a Federated User Preference Modeling (FUPM) framework for PPCDR. This framework first learns comprehensive user preferences and then privately transfers them to mitigate data sparsity issues, as shown in Fig. \ref{motivation} (c). Specifically, to explore comprehensive user preferences, we first design a contrastive feature alignment module to align user/item ID and review text features using contrastive learning (CL). We then introduce a potential interest mining module to extract users' potential interests by merging positive item embeddings and potentially positive item embeddings using Gaussian interpolation methods. To privately and accurately transfer user preferences, we propose an innovative private preference transfer module. It first learns local prototypes by averaging user embeddings within a group. These prototypes represent the collective preferences of multiple users rather than any single individual, making it more difficult for an attacker to associate a prototype with a specific user. To further strengthen privacy protection, we apply the LDP technique to these prototypes. These differentially private prototypes are then uploaded to the global server, where an adaptive prototype aggregation strategy is employed to learn global prototypes. The global prototypes are finally sent back to local domains to refine user preferences through knowledge transfer via CL.

The contributions of this work are summarized as follows:
\begin{itemize}
    \item We propose a novel FUPM framework, that addresses the data-sparsity problem in CDR while protecting user privacy.
    \item In FUPM, we design a comprehensive preference exploration module with contrastive feature alignment and potential interest mining components, which learns users' comprehensive interests from review texts and potentially positive items.
    \item In FUPM, we design a private preference transfer module that learns and transfers differentially private local and global prototypes to protect user privacy.
    \item Extensive experiments are conducted across four CDR tasks using Amazon and Douban datasets. The experimental results demonstrate the superiority of FUPM.
\end{itemize}

\section{Related Work}
\subsection{Cross Domain Recommendation}
Cross-domain recommendation (CDR) aims to address the data-sparsity problem by transferring cross-domain knowledge \cite{zang2023survey, zhu2021crossdomain}. Based on application scenarios, CDR methods can be classified into two categories: single-target CDR and multi-target CDR. Single-target CDR methods \cite{li2024aiming,man2017crossdomain,kang2019semisupervised} focus on improving recommendation performance in one specific target domain using information from one or more auxiliary domains. CUT \cite{li2024aiming} aims to solve the negative transfer and data-sparsity problems by explicitly filtering users' collaborative information. EMCDR \cite{man2017crossdomain} designs a mapping function to effectively transfer user preferences from the source domain to the target domain. In contrast, multi-target CDR approaches \cite{lu2023contrastive,zhu2020graphical,xu2023neurala,guo2023disentangleda} aim to simultaneously improve recommendation performance across multiple domains. CL-DTCDR \cite{lu2023contrastive} proposes a CL framework to address the data-sparsity problem. GA-DTCDR \cite{zhu2020graphical} introduces a graphical and attentional framework to learn comprehensive user
and item embeddings. DR-MTCDR \cite{guo2023disentangleda} disentangles user and item representations into domain-shared and domain-specific components and transfers the domain-shared information across domains.

Our method belongs to multi-target CDR. Although the above CDR methods have achieved notable improvements, they assume that user-item interaction data is shared across domains, leading to user privacy leakage. To address this problem, we introduce prototypes, LDP, and FL techniques to ensure data privacy.

\subsection{Privacy-Preserving Cross Domain Recommendation}
With the enactment of privacy protection laws and the increasing focus on user privacy \cite{7423759}, many scholars have begun studying PPCDR methods. These approaches leverage technologies, such as encryption \cite{wang2021poi}, LDP \cite{chen2022differential,chen2023winwin,gao2019privacypreserving}, and FL \cite{tian2024privacypreserving,liu2023federateda,yan2022fedcdr,liu2021fedct}, to protect users' sensitive information when transferring knowledge across domains. NATR \cite{gao2019crossdomain} first transfers user-irrelevant information across domains to protect user behavior data. PriCDR \cite{chen2022differential} then utilizes Differential Privacy (DP) to publish the rating matrix of the source domain, while P2FCDR \cite{chen2023winwin} utilizes the LDP technique to add noise to the transformed embeddings before transfer. PPGenCDR \cite{liao2023ppgencdr} further leverages adversarial methods to generate fake ratings for transfer to the target domain. Meanwhile, FPPDM \cite{liu2023federateda} and FedCDR \cite{yan2022fedcdr} introduce the FL framework to protect user privacy through distributed learning.

Although these PPCDR methods have achieved significant success in protecting privacy, they still exhibit limited privacy-preserving ability and suboptimal model performance. In this paper, we propose the FUPM framework to address these challenges.

\section{Methods}
\begin{figure*}[!htb] %H为当前位置，!htb为忽略美学标准，htbp为浮动图形
\centering %图片居中
\vspace{-4mm}
\includegraphics[width=1\textwidth]{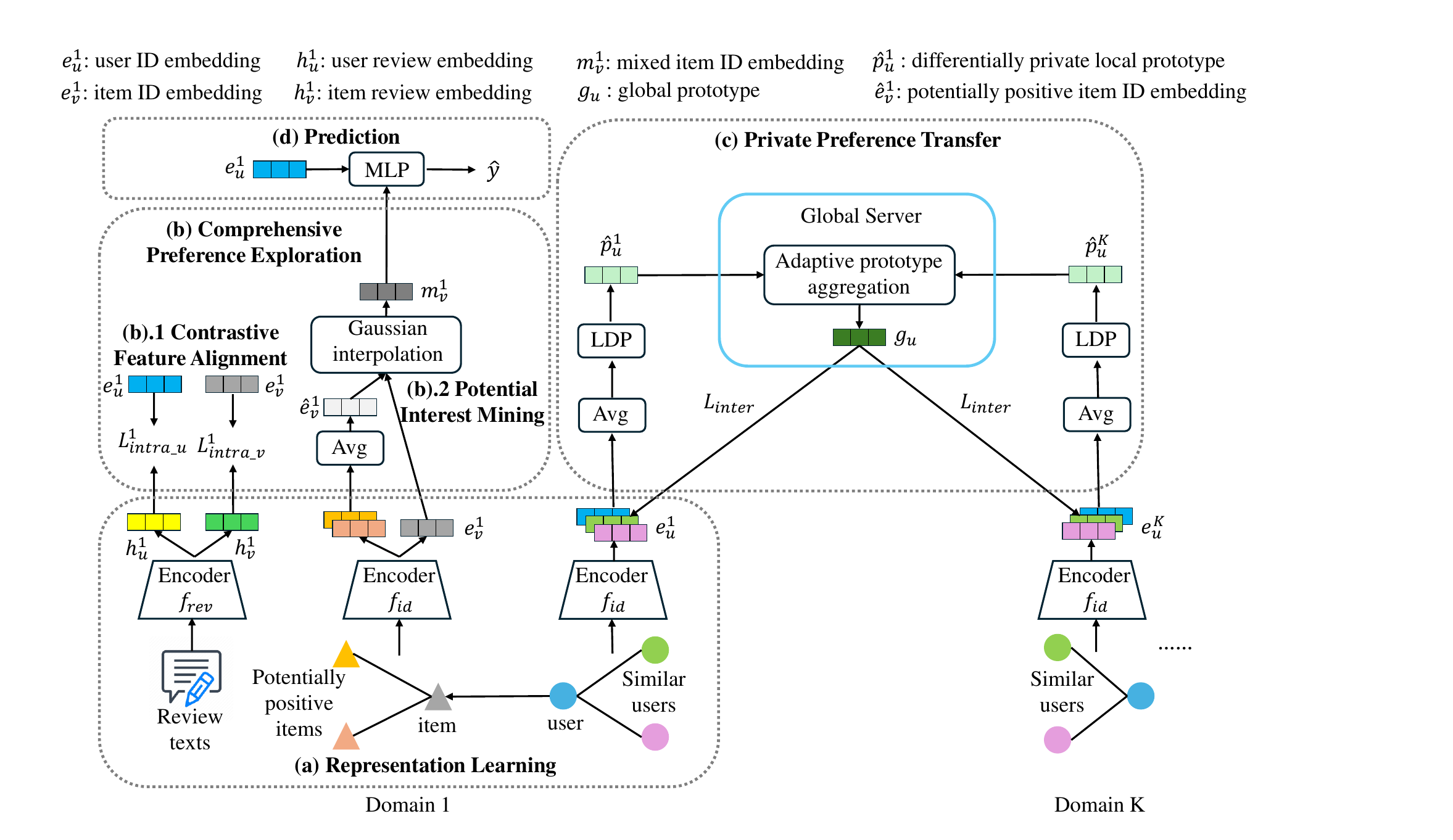}
\caption{The overall framework of FUPM.
It contains four modules: (a) \textbf{Representation Learning Module} aims to learn embeddings for user/item IDs and review texts. (b) \textbf{Comprehensive Preference Exploration Module} focuses on exploring comprehensive user preferences within each domain. It further divides into two components: (b).1 \textbf{Contrastive feature alignment} aims to align ID embeddings and review embeddings through CL. (b).2 \textbf{Potential Interest Mining} leverages potentially positive items to capture user's potential interests. (c) \textbf{Private Preference Transfer Module} aims to privately transfer user preferences across domains. (d) \textbf{Prediction Module} predicts user preferences.} %最终文档中希望显示的图片标题
\label{framework} %用于文内引用的标签
\vspace{-1mm}
\end{figure*}

We present the structure of our proposed framework, FUPM, in Fig. \ref{framework}. It contains four main modules: (1) \textbf{Representation Learning Module} aims to learn embeddings for user/item IDs and review texts; (2) \textbf{Comprehensive Preference Exploration Module} focuses on fully exploring user preferences by utilizing data within each domain. It is further divided into two components: 1) Contrastive Feature Alignment, which aligns ID embeddings and review text embeddings via CL. 2) Potential Interest Mining, which leverages potentially positive items to explore users' potential interests; (3) \textbf{Private Preference Transfer Module} aims to transfer user preferences across multiple domains without user privacy leakage; and (4) \textbf{Prediction Module} focuses on predicting user preferences towards items. Next, We illustrate the paradigm for domain $D^i$ as an example, and the corresponding paradigm for other domains can be easily inferred accordingly. We first provide the definitions and notations. Then, we introduce each module in detail.

% \begin{figure*}[!htb] %H为当前位置，!htb为忽略美学标准，htbp为浮动图形
% \centering %图片居中
% \vspace{-4mm}
% \includesvg[width=0.85\textwidth]{framework.svg}
% \caption{The overall framework of FUPM.
% It contains four modules: (a) \textbf{Representation Learning Module} aims to learn embeddings for user/item IDs and review texts. (b) \textbf{Comprehensive Preference Exploration Module} focuses on exploring comprehensive user preferences within each domain. It further divides into two components: (b).1 \textbf{Contrastive feature alignment} aims to align ID embeddings and review embeddings through CL. (b).2 \textbf{Potential Interest Mining} leverages potentially positive items to capture user's potential interests. (c) \textbf{Private Preference Transfer Module} aims to privately transfer user preferences across domains. (d) \textbf{Prediction Module} predicts user preferences.} %最终文档中希望显示的图片标题
% \label{framework} %用于文内引用的标签
% \vspace{-1mm}
% \end{figure*}

\subsection{Definitions and Notations}
In this paper, we focus on studying the privacy-preserving multi-target CDR scenario with overlapping users and distinct items. We assume that there are $K$ domains (clients) and a global server, where $D^i$ represents the $i-th$ domain. In each domain, $U$ represents the user set, $V^i$ denotes the item set, $R^i$ indicates the user-item review texts, and $Y^i\in\{0,1\}^{|U|\times|V^i|}$ is the user-item implicit interaction matrix, where each element represents whether the user has interacted with an item or not. 
% The mathematical notations are summarized
% in Table \ref{notations}. 
% \begin{table}[h]
% \centering
% \caption{Notations.}\label{notations}
% \begin{tabular}{c|c}
% \hline
% Symbols&Definitions and Notations\\
% \hline
% $*^i$& the i-th domain.\\
% K&the number of domains.\\
% U, V& user and item set\\
% R& review texts\\
% Y& user-item implicit interaction matrix.\\
% $\textbf{e}_u$, $\textbf{e}_v$&user and item ID embedding.\\
% $\textbf{h}_u$, $\textbf{h}_v$ & user and item review embedding.\\
% $\hat{\textbf{e}}_v^i$ & potential item embedding.\\
% $\textbf{m}_v^i$ & mixed item embedding.\\
% $\beta$ &the weight coefficient of the interpolation method.\\
% $\textbf{p}_u^i$ & the local prototype of user u.\\
% $\hat{\textbf{p}}_u^i$ & the protective local prototype of user u.\\
% $\alpha$ & the weight coefficient of intra-CL loss.\\
% $\gamma$ & the weight coefficient of inter-CL loss.\\
% \hline
% \end{tabular}
% \end{table}

\subsection{Representation Learning}
In this section, we utilize two encoders, $f_{id}$ and $f_{rev}$, to learn user/item ID embeddings and review embeddings respectively. The input of FUPM contains user and item identities as well as review texts. 

\textbf{ID embeddings:} We begin by transforming user and item identities into binarized sparse vectors via one-hot encoding. Then, we use an encoder $f_{id}$, which is an embedding layer, to project these sparse vectors into dense user ID embedding $\textbf{e}_u^i$ and item ID embedding $\textbf{e}_v^i$.

\textbf{Review embeddings:} First, we concatenate all review texts reviewed by a user into a single document, denoted as $D_u=\{d_1,d_2,...,d_m\}$. Similarly, for each item, we concatenate all its review texts from all users into a single document, denoted as $D_v=\{d_1,d_2,...,d_n\}$. We then employ document embedding models, i.e., Doc2vec \cite{le2014distributed} and Sentence Transformer \cite{reimers2019sentence}, to learn user review embedding $\textbf{h}_u^i$ and item review embedding $\textbf{h}_v^i$ respectively.

\subsection{Comprehensive Preference Exploration}
Most existing CDR methods primarily learn and transfer simple user preferences to alleviate data sparsity problems. However, they fail to utilize valuable side information to explore comprehensive user preferences, leading to suboptimal performance. In our study, we aim to learn comprehensive user interest from review texts and potentially positive items. 

\textbf{Contrastive Feature Alignment:} CL effectively addresses the issue of insufficient supervision signals by extracting valuable knowledge from abundant unlabeled data \cite{yu2023selfsupervised}. Inspired by this, we introduce CL to align ID embeddings and review embeddings, mitigating the sparsity problem in user-item interaction data.

 In the previous section, we learned the user/item ID embeddings $\textbf{e}_u^i$ and $\textbf{e}_v^i$, and review embeddings $\textbf{h}_u^i$ and $\textbf{h}_v^i$. In domain $D^i$, we consider the ID embedding and review embedding for the same user or item as a positive pair, and embeddings for different users or items as negative pairs in CL. For example, $(\textbf{e}_u^i, \textbf{h}_u^i)$ is a positive pair, while $(\textbf{e}_u^i, \textbf{h}_{u'}^i)$ is a negative pair where $u'\not =u$. The positive pair promotes consistency between the ID embedding and the review embedding for the same user, while the negative pair ensures that the embeddings for different users become more dissimilar. 

We use the widely-used InfoNCE \cite{he2020momentum} loss as the user intra-domain CL loss, which is defined as:
 \begin{equation}
\small
L_{intra\_u}^i = -log\frac{exp(f(\textbf{e}_u^i,\textbf{h}_u^i))}{exp(f(\textbf{e}_u^i,\textbf{h}_u^i))+\sum_{\textbf{h}_{u'}^i\in A(\textbf{h}_u^i),{u'}\not = u}exp(f(\textbf{e}_u^i,\textbf{h}_{u'}^i))},
\label{L_cl_u}
\end{equation}
where $A(\textbf{h}_u^i)$ represents the user review embedding set that doesn't include $\textbf{h}_u^i$. $f$ denotes the similarity function, shown as:
\begin{equation}
f(\textbf{e}_u^i,\textbf{h}_u^i) = \frac{\textbf{e}_u^i\cdot \textbf{h}_u^i}{||\textbf{e}_u^i||\times||\textbf{h}_u^i||}/\tau,
\end{equation}
where $\tau$ is the temperature coefficient, controlling the concentration strength of representation \cite{wang2021understanding}. Analogously, we can obtain the item intra-domain CL loss $L_{intra\_v}^i$.

\textbf{Potential Interest Mining:} Many existing approaches use items that users have interacted with as positive samples and non-interacted items as negative samples to learn user preferences during model training. However, non-interacted items do not necessarily indicate that users dislike them \cite{10462524}. They may be items that the recommendation system has not exposed to users. Therefore, it is worthwhile to identify items that users have not interacted with but might be interested in, which we regard as potentially positive items.

First, we calculate the similarity based on item review embeddings to identify potentially positive items. Specifically, for each user-item interaction pair $(u,v)$, we calculate the similarity score between the item $v$ and other items that the user $u$ has not interacted with before. We then select items whose similarity scores exceed a predefined threshold as potentially positive items for user $u$. Here, we empirically set the threshold to 0.5.

After identifying the potentially positive items, we average their ID embeddings to learn the potentially positive item ID embedding $\hat{\textbf{e}}_v^i$. Subsequently, we obtain the mixed item ID embedding $\textbf{m}_v^i$ by leveraging a linear interpolation method to integrate the item ID embedding $\textbf{e}_v^i$ and the potentially positive item ID embedding $\hat{\textbf{e}}_v^i$:
\begin{equation}
    \textbf{m}_v^i = \beta \textbf{e}_v^i + (1-\beta)\hat{\textbf{e}}_v^i,\ \beta \in (0,1), 
\end{equation}
where $\beta$ is sampled from a Gaussian distribution $Gaus(\mu,\delta)$. The advantages of utilizing this kind of interpolation method are as follows: (1) It creates a more comprehensive item ID embedding $\textbf{m}_v^i$ by interpolating $\textbf{e}_v^i$ and $\hat{\textbf{e}}_v^i$, which captures both explicit interactions and potential interests. 
(2) Using a random coefficient generated from the Gaussian distribution to perform interpolation, rather than a fixed coefficient, can generate diverse item embeddings, thereby providing richer information.

For negative items, we randomly sample several non-interacted items for interpolation, increasing the probability of including at least one true negative item. This process can reduce bias during the optimization of user preferences.

\subsection{Private Preference Transfer}
Recently, SOTA PPCDR methods have focused on transferring differentially private user-item interaction matrices or embeddings to protect data privacy. However, these methods offer limited privacy-preserving capabilities. To address this challenge, we store data locally within each domain using a decentralized FL framework. We then learn local prototypes and apply the LDP technique to these prototypes to further enhance user privacy protection.

First, we derive the local prototypes of all users. Specifically, for a given user $u$, we define a group as the set of users who share similar interests with $u$. We identify these similar users through the user-item interaction graph, focusing on users who have identical interactions with user $u$. We then compute the average of these users' ID embeddings to obtain the local prototype $\textbf{p}_u^i$ of user $u$:
\begin{equation}
    \textbf{p}_u^i = \frac{1}{N}\sum_{j\in N}\textbf{e}_j^i,
    \label{local_proto}
\end{equation}
where $N$ denotes the number of users in a group.

The aforementioned local prototype calculation with averaged embeddings can protect data privacy to a certain extent. These prototypes are generalized representations of user behavior, aggregating information from multiple users into a single, abstracted entity, making it challenging for attackers to infer sensitive information about individual users. To further enhance user privacy, we apply LDP techniques \cite{guaranteeing} to the local prototypes before transferring them. LDP ensures that the leakage of private information is bounded by a privacy budget, which is implemented using a random mechanism $M(\cdot)$:
\begin{equation}
\hat{\textbf{p}}_u^i=M(\textbf{p}_u^i) = clip(\textbf{p}_u^i,C)+Lap(0,\eta),
\label{ldp_effect}
\end{equation}
where $C$ is the clipping threshold that limits the value of $\textbf{p}_u^i$ to the scale of $C$. The purpose of clipping is to limit the sensitivity of individual data points, thereby controlling the amount of noise required to ensure LDP. Motivated by \cite{tian2024privacypreserving}, we use the $L_2$ norm for clipping. The parameter $\eta$ is the standard deviation of the Laplace distribution, which determines the strength of noise added. As demonstrated in \cite{wu2022fedgnn}, the upper bound of the privacy budget $\epsilon$ is given by $\frac{2C}{\eta}$. A smaller privacy budget improves privacy protection but sacrifices the model's performance. In the experimental section, we will provide an empirical study on the utility-privacy trade-off of LDP by analyzing the impact of the values $C$ and $\eta$.

Second, we upload the differentially private local prototype set $\hat{P}^i=\{\hat{\textbf{p}}_1^i,...,\hat{\textbf{p}}_u^i,...,\hat{\textbf{p}}_{|U|}^i\}$ from each domain to the global server to learn the global prototypes. For a given user $u$, we adopt an adaptive prototype aggregation strategy to model the global prototype:
\begin{equation}
    \textbf{g}_u = \sum_{i=1}^K \textbf{w}^i \hat{\textbf{p}}_u^i,
    \label{global_proto}
\end{equation}
where $i$ represents the i-th domain. $\textbf{w}^i$ is the weight of the differentially private local prototype $\hat{\textbf{p}}_u^i$ for user $u$, which is calculated as:
\begin{equation}
    \textbf{w}^i = \frac{|M^i|}{|M|}, \ \sum_{i=1}^K \textbf{w}^i = 1,
\end{equation}
where $|M^i|$ is the number of interactions of user $u$ in domain $D_i$, and $|M|$ is the total number of interactions of user $u$ across all domains. 

Finally, the global server sends back the global prototype set $G=\{\textbf{g}_1,\textbf{g}_2,...,\textbf{g}_{|U|},\}$ to the local domains.

To effectively transfer user preferences, we construct another CL task. Specifically, we consider the user ID embedding $\textbf{e}_u^i$ and the corresponding global prototype $\textbf{g}_u$ as the positive pair, and the global prototypes of different users as the negative pair. The inter-domain CL loss is defined as follows:
\begin{equation}
\small
L_{inter}^i = -log\frac{exp(f(\textbf{e}_u^i,\textbf{g}_u))}{exp(f(\textbf{e}_u^i,\textbf{g}_u))+\sum_{\textbf{g}_{u'}\in B(\textbf{g}_u),{u'}\not = u}exp(f(\textbf{e}_u^i,\textbf{g}_{u'}))},
\label{L_inter}
\end{equation}
where $B(\textbf{g}_u)$ is the global prototype set excluding $\textbf{g}_u$.

\subsection{Prediction}
After learning the comprehensive user ID embedding $\textbf{e}_u^i$, which represents comprehensive user preferences, and the mixed item ID embedding $\textbf{m}_v^i$, we input them into an MLP network to predict the probability that user $u$ will like item $v$. We employ multi-task learning to jointly optimize the following loss function: 
\begin{equation}
L^i = L_{prd}^i+\gamma(L_{intra\_u}^i+L_{intra\_v}^i)+\alpha L_{inter}^i,
\label{total_loss}
\end{equation}
where $L_{prd}^i$ represents the cross-entropy loss, and $\gamma$ and $\alpha$ are hyper-parameters to control the strengths of intra-CL and inter-CL, respectively. The overall learning algorithm is shown in Algorithm \ref{alg1}.

\begin{algorithm}
	%\textsl{}\setstretch{1.8}
\renewcommand{\algorithmicrequire}{\textbf{Input:}}
	\renewcommand{\algorithmicensure}{\textbf{Output:}}
	\caption{FUPM}
	\label{alg1}
	\begin{algorithmic}[1]
        \REQUIRE $D^i, \Theta^i, i=1,2,..,K$
        \STATE \textbf{Server executes:}
		\STATE Initialize global prototype set $G$.
		
		\FOR{round $r=1$ to $R$}
          \FOR{domain $i=1$ to $K$}
           \STATE $\hat{P}^i\gets LocalUpdate(i,G)$
          \ENDFOR
          \STATE Update global prototype set $G$ by Eq. (\ref{global_proto}).
        \ENDFOR
        \STATE \textbf{LocalUpdate($i,G$)}
		\FOR{epoch $e=1$ to $E$}
          \FOR{batch $b=1$ to $B$}
            \STATE Compute $L^i$ and update $\Theta^i$ by Eq. (\ref{total_loss}).
          \ENDFOR
        \ENDFOR
        \STATE Calculate the local prototype set $P^i$ by Eq. (\ref{local_proto}).
        \STATE Obtain the differentially private local prototype set $\hat{P}^i$ with LDP by Eq. (\ref{ldp_effect}).
        \RETURN $\hat{P}^i$
	\end{algorithmic}  
\end{algorithm}

\subsection{Model Analysis}
\subsubsection{Privacy Preserving Analysis}
The proposed FUPM effectively ensures user privacy through multiple mechanisms. Firstly, within the FL framework, data in each domain remains localized and is never shared with other domains, significantly reducing the risk of user privacy leakage \cite{wu2022fedgnn,meihan2022fedcdr}. Secondly, knowledge transfer across domains is facilitated through prototypes, which inherently protect data privacy \cite{tan2022fedproto}. These prototypes are one-dimensional vectors derived from averaging low-dimensional representations of samples within the same group, making the process irreversible. Lastly, we incorporate LDP on local prototypes before transferring them, ensuring that the leakage of user privacy is bounded and further strengthening privacy protection \cite{qi2020privacypreservinga}.
% For example, suppose we have a group of similar user embeddings $\{u_1:[0.3,0.4,0.2],u_2:[0.4,0.5,0.1],u_3:[0.3,0.1,0.7]\}$. The prototype is calculated by averaging these three user embeddings, resulting in $p=[0.33,0.33,0.33]$. If an attacker intercepts this prototype,  Without additional information, the attacker cannot determine the exact values of the individual embeddings of User $u_1$, $u_2$, or $u_3$. Because the prototype can represent different combinations of user embeddings, such as $\{u_1:[0.25,0.3,0.1],u_2:[0.15,0.2,0.3],u_3:[0.6,0.5,0.6]\}$ or $\{u_1:[0.1,0.5,0.3],u_2:[0.4,0.2,0.6],u_3:[0.5,0.3,0.1]\}$. The number of possible combinations that can result in the same prototype increases exponentially with the number of users and the dimensions of the embeddings. This combinatorial complexity further enhances privacy.

\subsubsection{Extension to Partial User Overlap} In this paper, we study a PPCDR scenario with overlapping users but non-overlapping items across domains. In real-world applications, partial user overlap is common. Therefore, we extend FUPM to handle partial user overlap. Specifically, we first utilize all user data, not just the overlapping users, to learn local prototypes. For user preference transfer, we divide the process into two parts: (1) For overlapping users, we transfer knowledge through global prototypes, as in FUPM. (2) For non-overlapping users, we rely on similar overlapping users within the domain to transfer knowledge since we cannot bridge different domains due to privacy protection regulations. Specifically, for each non-overlapping user, we first find similar overlapping users by calculating local prototype similarity. Then, we transfer the user preferences of these similar overlapping users.

\section{Experiments}
To evaluate the performance of FUPM, we conduct a series of extensive experiments on widely-used Amazon and Douban datasets to answer the following research questions:
\begin{itemize}
\item RQ1: How does FUPM perform compared to SOTA baselines?
\item RQ2: How do different components of FUPM affect its recommendation performance?
\item RQ3: Does FUPM effectively address the PPCDR problem, particularly in terms of data sparsity and privacy protection?
\item RQ4: How do key hyper-parameters influence the recommendation performance of FUPM?
\end{itemize}

\subsection{Experimental Settings}
\subsubsection{Datasets}
Inspired by SOTA CDR methods \cite{liu2020cross, liu2023federateda, chen2022differential, zhao2024crossdomain}, we evaluate the performance of FUPM using the widely-used Amazon\footnote{https://cseweb.ucsd.edu/~jmcauley/datasets/amazon/links.html} and Douban\footnote{https://www.dropbox.com/s/u2ejjezjk08lz1o/Douban.tar.gz?e=2\&dl=0} datasets. Specifically, we select three subsets - Phone, Sport, and Cloth - from the Amazon dataset, and three subsets - Book, Movie, and Music - from the Douban dataset to construct four CDR tasks. The detailed statistics of these datasets are shown in Table \ref{dataset_statistic}. For all datasets, we convert explicit ratings into implicit binary values of 0 and 1, representing whether a user has interacted with an item. During the training process, for each user, we randomly select one item that the user has not interacted with before as the negative sample and take the corresponding interacted item as the positive sample. To improve the data quality, we filter out users with fewer than 5 interactions and items with less than 10 interactions by following a general practice~\cite{chen2023winwin}. We utilize Doc2Vec to learn user/item review embeddings for the Amazon dataset, leveraging its proficiency in English text processing. On the other hand, we use the Sentence Transformer to learn user/item review embeddings for the Douban dataset, capitalizing on its adeptness in handling Chinese text.

\begin{table}[h]
\centering
\footnotesize
\caption{Statistics of the datasets for four CDR tasks.}\label{dataset_statistic}
\begin{tabular}{c c c c c c}
\hline
Tasks& Datasets & \#Users & \#Items & \#Ratings &  Density\\
\hline
\multirow{2}{*}{Task 1}&Phone & 4,998&	14,618&	47,444&	0.0649\%\\
&Sport & 4,998&	22,101&	55,556&	0.0503\%\\
\hline
\multirow{2}{*}{Task 2}&Book & 909&	3,222&	46,374&	1.583\%\\
&Music & 909&	2,546&	40,761&	1.761\%\\
\hline
\multirow{3}{*}{Task 3}&Sport & 2,229&	15,321&	30,512&	0.0893\%\\
&Cloth & 2,229&	16,664&	25,660&	0.0691\%\\
&Phone&2,229&	9,789&	25,175&	0.1154\%\\
\hline
\multirow{3}{*}{Task 4}&Book & 901&	3,222&	46,200&	1.591\%\\
&Movie & 901&	14,501&	560,675&	4.291\%\\
&Music & 901&	2,546&	40,512&	1.767\%\\
\hline
\end{tabular}
\end{table}

\subsubsection{Evaluation metrics}
We use the leave-one-out method, widely employed in literature \cite{he2017neural, liu2020cross, zhu2023domain}, to evaluate the performance of FUPM. Specifically, for each user-item interaction pair in the test set, we randomly select 99 items that the user has not interacted with before as negative samples. We then use the FUPM model to predict scores for these 100 items for the ranking process. Hit Ratio (HR) and Normalized Discounted Cumulative Gain (NDCG) are used as evaluation metrics, 
%and Mean Reciprocal Rank (MRR),
which are  commonly used in existing work on CDR \cite{li2020ddtcdr, guo2023disentangleda, zhu2023unified}. We set the length of the recommendation list, denoted as N, to 5 and 10, respectively, for evaluations. To minimize randomness, we conduct five training and testing cycles for each CDR task and report the average results. 

\subsubsection{Comparison methods}
To validate the superior performance of FUPM, we compare it against three sets of representative baselines widely used for CDR tasks \cite{lu2023contrastive, chen2023winwin,zhu2023unified}: 

\noindent \textbf{Single-Domain Recommendation}
\begin{itemize}
    \item \textbf{NeuMF} \cite{he2017neural} combines neural networks with matrix factorization techniques to capture complex user-item interaction patterns.
    \item \textbf{SSL} \cite{yao2021selfsupervised} utilizes self-supervised learning to effectively learn item features, addressing the label sparsity problem.
    \item \textbf{FedNCF} \cite{perifanis2022federated} introduces FL into the neural collaborative learning model to protect user’s privacy.
\end{itemize}
\textbf{Cross-Domain Recommendation}
\begin{itemize}
    \item \textbf{BiTGCF} \cite{liu2020cross} is a graph collaborative filtering-based CDR model that conducts bidirectional high-order information transfer to enhance the model performance.
    \item \textbf{DDTCDR} \cite{li2020ddtcdr} is a dual-learning CDR method designed to learn an orthogonal mapping function to transfer user preferences across domains.
    \item \textbf{GA-DTCDR} \cite{zhu2020graphical} is a graphical and attentional CDR framework that learns representative embeddings to improve recommendation accuracy.
    \item \textbf{CL-DTCDR} \cite{lu2023contrastive} is designed to solve data sparsity and sparse overlapping user problems using two CL tasks.
    \item \textbf{GA-MTCDR-P} \cite{zhu2023unified} is an enhanced model of GA-DTCDR, incorporating graph neural networks and multi-domain attention mechanisms to address the problem of negative transfer.
\end{itemize}
\textbf{Privacy-Preserving CDR}
\begin{itemize}
    \item \textbf{PriCDR} \cite{chen2022differential} utilizes LDP technology to publish the rating matrix in the source domain and subsequently transfers the published matrix to the target domain.
    \item \textbf{P2FCDR} \cite{chen2023winwin} is a privacy-preserving federated CDR framework that uses LDP techniques to protect user embeddings when transferring cross-domain knowledge.
\end{itemize}

\subsubsection{Parameter Settings}
We implemented FUPM in Python using the PyTorch framework. 
%All baselines are executed using their GitHub source code, with hyperparameters carefully tuned. 
The optimal hyper-parameters are determined by optimizing the loss function \eqref{total_loss} using the Adam optimizer \cite{KingBa15} with a learning rate of 0.001. The embedding dimension is set to 256, the batch size to 128, and epochs to 200. To balance recommendation performance and CL tasks, we set $\alpha$ to 0.05 and $\gamma$ to 0.2. Additionally, we employ early stopping and normalization techniques to prevent overfitting. All baselines are executed using their GitHub source code, with hyper-parameters carefully tuned on our datasets for the best performance.

\subsection{Experimental Results (for RQ1)}
To address RQ1, we evaluate FUPM and baselines using HR@\{5,10\} and NDCG@\{5,10\} on four CDR tasks. Table \ref{experimental_results} presents the results. 
%with the best performance in bold and the second-best underlined. 
We can summarize that:
\begin{itemize}
    \item \textbf{For single-domain methods}: (1) The CDR approaches perform better than single-domain methods. This shows that cross-domain knowledge can improve recommendation performance. (2) Compared to NeuMF, SSL achieves superior performance, highlighting the important role of CL in improving model effectiveness. (3) FedNCF, a federated collaborative filtering method, performs worse than NeuMF due to the decentralized interaction data. This shows that federated methods can protect user privacy.
    \item \textbf{For CDR methods}: (1) The performance of different CDR methods demonstrates that the choice of transfer strategies significantly affects recommendation performance. (2) GA-DTCDR outperforms BiTGCF, despite both being GNN-based methods. This superiority is due to the attention mechanism used in GA-DTCDR, which helps avoid the negative transfer problem. (3) CL-DTCDR achieves the best performance in most cases compared to other CDR models, demonstrating the effectiveness of CL for cross-domain knowledge transfer. (4) GA-MTCDR-P performs better than GA-DTCDR because it incorporates more domains (datasets). (5) P2FCDR outperforms PriCDR, indicating that dual-target CDR methods exceed single-target ones.
    % (5) Compared to CL-DTCDR, PriCDR performs worse, which shows that the use of local differential privacy technique might reduce model performance.
    \item \textbf{For our FUPM}: (1) Among all CDR methods, FUPM consistently exceeds other baselines across all CDR tasks. This superiority illustrates that our method, which learns comprehensive user preferences, can effectively solve the data-sparsity problem. Although introducing privacy-preserving techniques may decrease model performance, FUPM still outperforms other CDR methods. Specifically, FUPM improves over the leading baseline CL-DTCDR by an average of 13.28\% in HR@10 and 11.97\% in NDCG@10. Additionally, CL-DTCDR only transfers knowledge between two domains, while our method handles more than two domains, as seen in Tasks 3 and 4. (2) Compared with PPCDR methods, i.e., PriCDR and P2FCDR, our model obtains the best performance. It shows that effectively exploring comprehensive user preferences can improve recommendation accuracy.
\end{itemize}
\begin{table*}[htbp]
\centering
\caption{Experimental Results on four CDR tasks. The best performance is in bold, and the second best is underlined. The superscript ``*" indicates the statistical
significance for $p < 0.01$ compared to the best baseline.}
\label{experimental_results}
\fontsize{7.5}{8.5}\selectfont
\setlength{\tabcolsep}{3.5pt}
\begin{tabular}{c c c c c c c c c c c c c c c c c}
\hline
\multirow{3}{*}{Datasets} & \multicolumn{2}{c}{\multirow{3}{*}{Metrics}} & \multicolumn{3}{c}{Single domain methods} & \multicolumn{5}{c}{CDR methods}& \multicolumn{2}{c}{PPCDR methods}&\multicolumn{2}{c}{Ours}\\ 
% \cline{4-6}\cline{7-10}\cline{11-12}\cline{13-14}
\cmidrule(lr){4-6}\cmidrule(lr){7-11}\cmidrule(lr){12-13}\cmidrule(lr){14-15}
   & &&NeuMF & SSL & FedNCF & BiTGCF & DDTCDR &GA-DTCDR&CL-DTCDR& GA-MTCDR-P&PriCDR&P2FCDR&FUPM&Imp \\ \hline
\multirow{4}{*}{Phone} &\multirow{2}{*}{N=5}& HR &0.2779&0.2802&0.2583&0.3269&0.3260&0.3317&\underline{0.4728}&0.3367&0.3513&0.3899&\textbf{0.6131}*&14.03\%\\  
&& NDCG &0.1932&0.2098&0.1905&0.2378&0.2038&0.2420&\underline{0.3260}&0.2431&0.2670&0.3047&\textbf{0.4955}*&16.95\%  \\
% && MRR &0.1654&0.1694&0.1547&0.2083&0.1639&0.2127&0.2976&0.2135&0.2525&0.2766&0.4566&15.90\%  \\ 
&\multirow{2}{*}{N=10}& HR &0.3697&0.3724&0.3173&0.4188&0.4992&0.4679&\underline{0.6198}&0.4690&0.4283&0.4896&\textbf{0.7291}*&10.93\%  \\  
&& NDCG &0.2231&0.2567&0.2217&0.2671&0.2596&0.2857&\underline{0.3833}&0.2885&0.2962&0.3369&\textbf{0.5332}*&14.99\%  \\
% && MRR &0.1778&0.2031&0.1769&0.2203&0.1868&0.2306&0.3269&0.2341&0.2458&0.2899&0.4722&14.53\%\\ 
\cline{2-15}
\multirow{4}{*}{Sport} &\multirow{2}{*}{N=5}& HR &0.1955&0.2044&0.1917&0.2763&0.3695&0.2969&\underline{0.5020}&0.3004&0.3249&0.3135&\textbf{0.6010}*&9.90\%  \\  
&& NDCG &0.1351&0.1416&0.1245&0.2074&0.2558&0.2362&\underline{0.3687}&0.2394&0.2441&0.2494&\textbf{0.4855}*&11.68\%  \\
% && MRR &0.1153&0.1257&0.1058&0.1846&0.2182&0.2096&0.3281&0.2120&0.2105&0.2282&0.4472&11.91\%  \\ 
&\multirow{2}{*}{N=10}& HR &0.2719&0.2863&0.2707&0.3631&0.4768&0.3901&\underline{0.6278}&0.3941&0.3925&0.3827&\textbf{0.7195}*&9.17\%  \\  
&& NDCG &0.1598&0.1702&0.1473&0.2352&0.2906&0.2662&\underline{0.4025}&0.2682&0.2735&0.2719&\textbf{0.5238}*&12.13\%  \\
% && MRR &0.1254&0.1502&0.1169&0.1959&0.2326&0.2120&0.3460&0.2157&0.2479&0.2376&0.4630&11.70\%\\
\hline
\multirow{4}{*}{Book} &\multirow{2}{*}{N=5}& HR &0.1705&0.1817&0.1650&0.2860&0.2677&0.2943&0.2852&0.2931&0.2049&\underline{0.2958}&\textbf{0.3677}*&7.19\%  \\  
&& NDCG &0.1192&0.1410&0.1047&0.1931&0.2001&0.2115&0.2112&0.2134&0.1632&\underline{0.2148}&\textbf{0.2591}*&4.43\%  \\
% && MRR &0.1024&0.1243&0.1012&0.1627&0.1782&0.1910&0.1969&0.1893&0.1364&0.1950&0.2233&2.64\%  \\ 
&\multirow{2}{*}{N=10}& HR & 0.2607&0.2789&0.2584&0.4224&0.4294&0.4455&0.4308&\underline{0.4463}&0.3130&0.4080&\textbf{0.5188}*&7.25\% \\  
&& NDCG &0.1479&0.1588&0.1358&0.2369&0.2516&0.2541&\underline{0.2649}&0.2561&0.2433&0.2511&\textbf{0.3076}*&4.27\%  \\
% && MRR &0.1139&0.1312&0.1049&0.1806&0.1990&0.2043&0.2006&0.2051&0.2101&0.2199&0.2432&2.33\%\\ 
\cline{2-15}
\multirow{4}{*}{Music} &\multirow{2}{*}{N=5}& HR &0.1265&0.1483&0.1205&0.2070&0.2150&0.2248&0.2225&\underline{0.2284}&0.1987&0.2208&\textbf{0.2374}*&0.90\%  \\  
&& NDCG &0.0830&0.0918&0.0836&0.1263&0.1318&0.1469&\underline{0.1475}&0.1458&0.1414&0.1397&\textbf{0.1521}*&0.46\%  \\
% && MRR &0.0689&0.0732&0.0639&0.0931&0.1013&0.1148&0.1165&0.1164&0.1059&0.1099&0.1242&0.77\%  \\ 
&\multirow{2}{*}{N=10}& HR &0.2211&0.2344&0.2173&0.3188&0.3250&0.3359&\underline{0.3536}&0.3486&0.3024&0.3463&\textbf{0.3630}*&0.94\%  \\  
&& NDCG &0.1132&0.1498&0.1093&0.1549&0.1628&0.1827&0.1833&\underline{0.1851}&0.1635&0.1772&\textbf{0.1928}*&0.77\%  \\
% && MRR &0.0812&0.0906&0.0740&0.1229&0.1219&0.1295&0.1312&0.1273&0.1238&0.1254&0.1410&0.98\%\\
\hline
\multirow{4}{*}{Sport} &\multirow{2}{*}{N=5}& HR & 0.1772&0.2444&0.1658&0.2369&0.3491&0.2497&\underline{0.4074}&0.2506&0.2584&0.2597&\textbf{0.6120}*&20.46\% \\  
&& NDCG &0.1181&0.1265&0.1046&0.1703&0.2461&0.1852&\underline{0.2847}&0.1874&0.1960&0.2123&\textbf{0.4880}*&20.33\%  \\
% && MRR &0.0987&0.1207&0.0910&0.1483&0.2122&0.1508&0.2444&0.1539&0.1754&0.1966&0.4470&20.26\%  \\ 
&\multirow{2}{*}{N=10}& HR &0.2562&0.3005&0.2461&0.3185&0.4585&0.3299&\underline{0.5500}&0.3341&0.3410&0.3266&\textbf{0.7354}*&18.54\%  \\  
&& NDCG & 0.1437&0.2444&0.1357&0.1967&0.2815&0.2045&\underline{0.3307}&0.2073&0.2163&0.2338&\textbf{0.5280}*&19.73\% \\
% && MRR &0.1094&0.228&0.0925&0.1591&0.2268&0.1630&0.2633&0.1652&0.1782&0.2054&0.4636&20.03\%\\ 
\cline{2-15}
\multirow{4}{*}{Cloth} &\multirow{2}{*}{N=5}& HR & 0.1274&0.1529&0.1168&0.3472&0.3204&0.3531&\underline{0.4029}&0.3562&0.2898&0.1956&\textbf{0.5856}*&18.27\% \\  
&& NDCG &0.0876&0.1398&0.0762&0.2006&0.2830&0.2023&\underline{0.3836}&0.2047&0.2121&0.1671&\textbf{0.4625}*&7.89\%  \\
% && MRR & 0.0746&0.1237&0.0625&0.1853&0.2307&0.1988&0.2642&0.2018&0.1863&0.1578&0.4219&15.77\% \\ 
&\multirow{2}{*}{N=10}& HR &0.1943&0.1974&0.1835&0.4221&0.4797&0.4337&\underline{0.5428}&0.4369&0.3715&0.2516&\textbf{0.7201}*&17.73\%  \\  
&& NDCG & 0.1092&0.1450&0.0914&0.3248&0.3020&0.3313&0.3285&\underline{0.3341}&0.2092&0.1850&\textbf{0.5059}*&17.18\% \\
% && MRR &0.0835&0.1295&0.0732&0.2953&0.2784&0.2964&0.2626&0.2998&0.1595&0.1650&0.4397&13.99\%\\
\cline{2-15}
\multirow{4}{*}{Phone} &\multirow{2}{*}{N=5}& HR &0.2364&0.2475&0.2267&0.2665&0.2985&0.2770&0.3374&0.2794&0.2885&\underline{0.3418}&\textbf{0.6113}*&26.95\%  \\  
&& NDCG &0.1600&0.1701&0.1520&0.1855&0.1752&0.1954&0.2302&0.1975&0.2078&\underline{0.2661}&\textbf{0.4696}*&20.35\%  \\
% && MRR &0.1349&0.1434&0.1269&0.1589&0.1352&0.1689&0.1950&0.1723&0.1809&0.2413&0.4228&18.15\%  \\ 
&\multirow{2}{*}{N=10}& HR & 0.3428&0.3488&0.3326&0.3701&0.4955&0.3856&\underline{0.4715}&0.3895&0.3490&0.4163&\textbf{0.7374}*&24.19\% \\  
&& NDCG &0.1943&0.2030&0.1825&0.2189&0.2388&0.2305&0.2733&0.2346&0.2188&\underline{0.2903}&\textbf{0.5104}*&22.01\%  \\
% && MRR &0.1490&0.1572&0.1360&0.1726&0.1614&0.1833&0.2126&0.1857&0.1778&0.2512&0.4396&18.84\%\\
\hline
\multirow{4}{*}{Book} &\multirow{2}{*}{N=5}& HR &0.1554&0.1633&0.1476&0.2219&\underline{0.2686}&0.2354&0.2408&0.2347&0.1687&0.2587&\textbf{0.3454}&7.68\%  \\  
&& NDCG & 0.1037&0.1326&0.0949&0.1997&0.1989&0.2009&0.1916&\underline{0.2012}&0.1115&0.1916&\textbf{0.2437}&4.25\% \\
% && MRR & 0.0868&0.1159&0.0821&0.1725&0.1763&0.1896&0.1790&0.1882&0.0930&0.1697&0.2083&1.87\% \\ 
&\multirow{2}{*}{N=10}& HR & 0.2531&0.2665&0.2461&0.3306&0.3320&0.3420&\underline{0.3529}&0.3457&0.2575&0.3486&\textbf{0.4875}&13.46\% \\  
&& NDCG &0.1348&0.1597&0.1256&0.2475&0.2511&0.2525&0.2472&0.2548&0.1384&\underline{0.2672}&\textbf{0.2879}&2.07\%  \\
% && MRR &0.0994&0.1329&0.0912&0.1421&0.1475&0.1686&0.1633&0.1695&0.1025&0.1701&0.2270&5.69\%\\ 
\cline{2-15}
\multirow{4}{*}{Movie} &\multirow{2}{*}{N=5}& HR &0.3996&0.4077&0.3874&0.3573&\underline{0.5018}&0.3607&0.4984&0.3624&0.3940&0.3984&\textbf{0.6584}*&15.66\%  \\  
&& NDCG & 0.2773&0.2874&0.2639&0.2249&0.3277&0.2543&\underline{0.3346}&0.2563&0.2530&0.2798&\textbf{0.5006}*&16.60\% \\
% && MRR & 0.2370&0.2408&0.2249&0.2009&0.2500&0.2191&0.2609&0.2219&0.2067&0.2407&0.4484&18.75\% \\ 
&\multirow{2}{*}{N=10}& HR & 0.5350&0.5477&0.5230&0.5304&\underline{0.6060}&0.5406&0.6049&0.5432&0.5549&0.5128&\textbf{0.7851}*&17.91\% \\  
&& NDCG &0.3211&0.3402&0.3126&0.3809&0.3277&0.3998&0.3385&\underline{0.4012}&0.3226&0.3169&\textbf{0.5418}*&14.06\%  \\
% && MRR &0.2551&0.2672&0.2439&0.2340&0.2407&0.2380&0.2431&0.2396&0.2509&0.2461&0.4655&21.46\%\\
\cline{2-15}
\multirow{4}{*}{Music} &\multirow{2}{*}{N=5}& HR & 0.1310&0.1411&0.1209&0.1963&0.2028&0.2010&\underline{0.2054}&0.2045&0.1287&0.1999&\textbf{0.2133}*&0.79\% \\  
&& NDCG &0.0850&0.0829&0.0736&0.0910&0.0950&0.1025&\underline{0.1242}&0.1037&0.0773&0.1107&\textbf{0.1387}*&1.45\%  \\
% && MRR &0.0700&0.0771&0.0723&0.0865&0.0831&0.0900&0.1042&0.1029&0.0606&0.1045&0.1144&0.99\%  \\ 
&\multirow{2}{*}{N=10}& HR & 0.2164&0.2243&0.2056&0.3206&0.2692&0.3309&0.2552&\underline{0.3316}&0.2031&0.3186&\textbf{0.3412}*&0.96\% \\  
&& NDCG & 0.1121&0.1397&0.1067&0.1635&0.1652&0.1518&\underline{0.1654}&0.1502&0.1012&0.1393&\textbf{0.1800}*&1.46\% \\
% && MRR &0.0809&0.1138&0.0762&0.1078&0.0936&0.1179&0.1096&0.1180&0.0708&0.1262&0.1314&0.52\%\\
\hline
\end{tabular}
\end{table*}

\subsection{In-depth Analysis} 
% In the previous section, we validated our model's superiority by comparing its performance with SOTA baselines. 
We demonstrate the effectiveness of FUPM through in-depth analysis.
\subsubsection{Ablation Studies (for RQ2)}
To evaluate the impact of various components on the overall recommendation performance, we conduct ablation studies across four CDR tasks. We create several variants to evaluate the effect of specific components:
(1) \textbf{w/o intra\_CL} removes the CL task in the contrastive feature alignment module.
(2) \textbf{intra\_sum} integrates ID and review embeddings using element-wise sum.
(3) \textbf{w/o inter\_CL} removes the CL task in the private preference transfer module, meaning that cross-domain knowledge is not transferred.
(4) \textbf{inter\_sum} uses element-wise sum instead of CL for knowledge transfer.
(5) \textbf{w/o PI} removes the potential interest mining module.
(6) \textbf{rand\_sam} identifies potentially positive items by randomly sampling from non-interacted items instead of selecting similar items.

Based on the experimental results in Table \ref{ablation_studies}, we observe that: (1) FUPM consistently outperforms both w/o \textbf{intra\_CL} and \textbf{intra\_sum}, demonstrating the effectiveness of CL in learning comprehensive user preferences, and improving model performance. (2) The performance of \textbf{w/o PI} significantly degrades compared to FUPM, highlighting the importance of the potential interest mining module in enhancing recommendation performance. (3) FUPM achieves better results than \textbf{rand\_sam}, indicating that identifying potentially positive items through similarity calculations can improve model performance. (4) Compared with \textbf{w/o inter\_CL} and \textbf{inter\_sum}, FUPM achieves the best results, showing that transferring cross-domain knowledge through CL improves recommendation accuracy.

\begin{table}[htbp]
\centering
\vspace{-3mm}
\fontsize{6.5}{7.5}\selectfont
\setlength{\tabcolsep}{3.0pt}
\caption{Ablation Studies.}
\label{ablation_studies}
\begin{tabular}{c c c c c c c c c}
\hline
\multirow{2}{*}{Datasets} & \multirow{2}{*}{Metrics} & \multicolumn{6}{c}{Model variants} & \multirow{2}{*}{Ours} \\ \cline{3-8}
 &  & \makecell{w/o \\inter\_CL} & inter\_sum & \makecell{w/o\\ intra\_CL} & intra\_sum & w/o PI &rand\_sam&  \\ \hline
\multirow{2}{*}{Phone} & HR &0.7233&0.6066&0.6757&0.7141&0.6145&0.5968&0.7291  \\  
& NDCG & 0.5233&0.4379&0.5060&0.4983&0.3931&0.3827&0.5332 \\
% & MRR &0.4711&0.3857&0.4538&0.4315&0.3244&0.3162&0.4722  \\ 
\cline{2-9}
\multirow{2}{*}{Sport} & HR &0.7123&0.5842&0.5972&0.7093&0.6086&0.5820&0.7195 \\  
& NDCG &0.5207&0.4101&0.4656&0.5039&0.3880&0.3698&0.5238  \\
% & MRR & 0.4531&0.3564&0.4248&0.4397&0.3196&0.3045&0.4630 \\ 
\hline
\multirow{2}{*}{Book} & HR &0.5127&0.4499&0.4775&0.5038&0.3795&0.3630&0.5188  \\  
& NDCG &0.3041&0.2632&0.2775&0.2891&0.2178&0.2011&0.3076  \\
% & MRR & 0.2407&0.2064&0.2166&0.2208&0.1684&0.1519&0.2432 \\ 
\cline{2-9}
\multirow{2}{*}{Music} & HR &0.3585&0.3520&0.3549&0.3432&0.3608&0.3212&0.3630 \\  
& NDCG & 0.1873&0.1833&0.1870&0.1734&0.1919&0.1737&0.1928 \\
% & MRR&0.1326&0.1321&0.1407&0.1221&0.1402&0.1291&0.1410  \\ 
\hline
\multirow{2}{*}{Sport} & HR & 0.5886&0.5958&0.5666&0.6981&0.5769&0.5428&0.7354 \\  
& NDCG &0.3697&0.4020&0.4179&0.4715&0.3558&0.3254&0.5280  \\
% & MRR &0.3021&0.3421&0.3722&0.4008&0.2877&0.2588&0.4636  \\ 
\cline{2-9}
\multirow{2}{*}{Cloth} & HR &0.5500&0.5092&0.4814&0.6411&0.5626&0.5173&0.7201  \\  
& NDCG & 0.3423&0.3246&0.3091&0.3924&0.3563&0.3172&0.5059 \\
% & MRR &0.2782&0.2677&0.2560&0.3157&0.2927&0.2558&0.4397  \\ 
\cline{2-9}
\multirow{2}{*}{Phone} & HR &0.5747&0.5689&0.6245&0.7088&0.5868&0.5639&0.7374 \\  
& NDCG &0.3588&0.3682&0.4491&0.4727&0.3602&0.3447&0.5104  \\
% & MRR &0.2926&0.3064&0.3952&0.3993&0.2905&0.2773&0.4396  \\ 
\hline
\multirow{2}{*}{Book} & HR &0.4839&0.3374&0.4506&0.4828&0.3529&0.3663&0.4875  \\  
& NDCG &0.2858&0.1917&0.2527&0.2711&0.1965&0.1982&0.2879  \\
% & MRR & 0.2253&0.1475&0.1923&0.2065&0.1495&0.1478&0.2270 \\ 
\cline{2-9}
\multirow{2}{*}{Movie} & HR & 0.7780&0.7458&0.7414&0.7558&0.6759&0.6448&0.7851 \\  
& NDCG &0.5353&0.5039&0.4888&0.4965&0.4180&0.4002&0.5418  \\
% & MRR &0.4564&0.4291&0.4106&0.4157&0.3393&0.3254&0.4655  \\ 
\cline{2-9}
\multirow{2}{*}{Music} & HR & 0.3329&0.2897&0.3318&0.3352&0.3374&0.3274&0.3412 \\  
& NDCG & 0.1765&0.1545&0.1792&0.1743&0.1754&0.1692&0.1800 \\
% & MRR &0.1232&0.1131&0.1298&0.1259&0.1233&0.1211&0.1314  \\ 
\hline
\end{tabular}
\vspace{-1mm}
\end{table}

\subsubsection{Interaction density M (for RQ3)} 
\begin{figure}[!htb]
\centering
\vspace{-5mm}
\subfloat[\scriptsize Book\&Music on Book]{\includegraphics[width=1.65in]{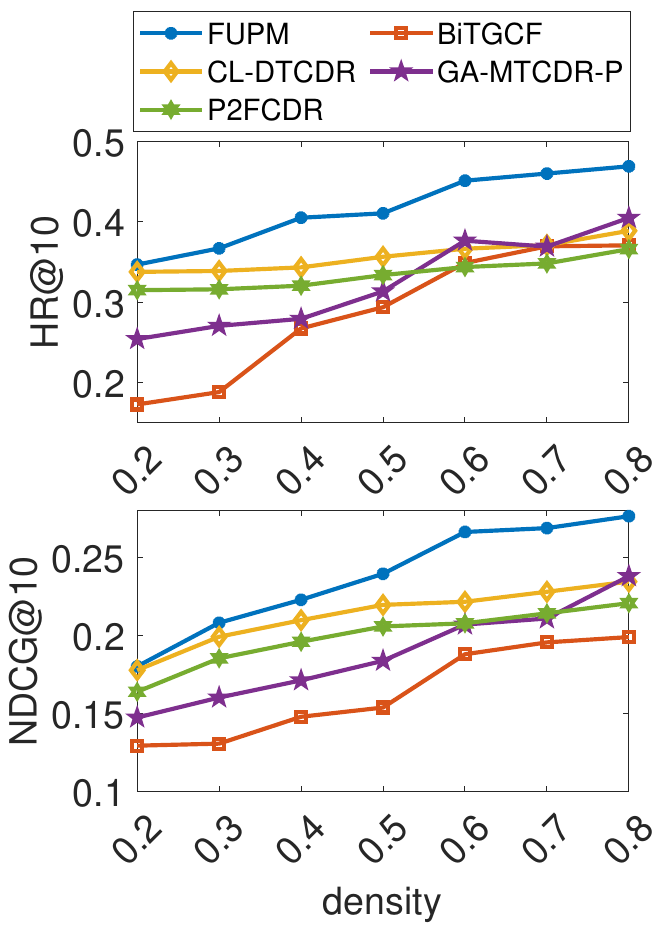}}
\hfil
\subfloat[\scriptsize Book\&Music on Music]{\includegraphics[width=1.65in]{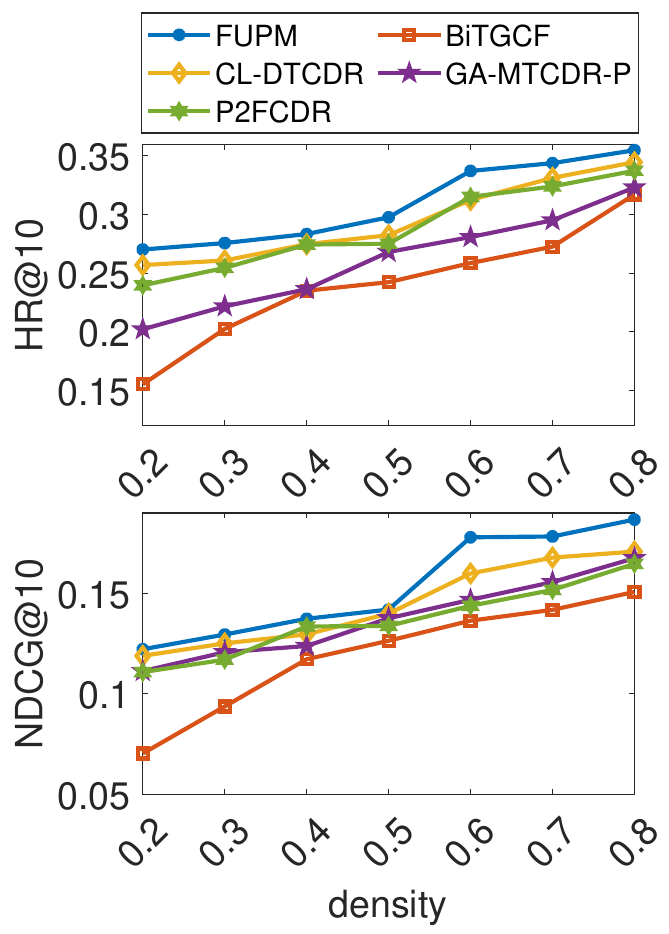}}
\hfil
\caption{The performance of different interaction densities.}
\label{inter_density}
\end{figure}

To evaluate whether FUPM can effectively address the data-sparsity problem, we conducted experiments on Tasks 1 and 2 with varying interaction densities. Specifically, we vary the number of interacted items for each user in the training set based on interaction densities ranging from 0.2 to 0.8. These varying densities illustrate different levels of sparsity, with a higher interaction density indicating lower sparsity. For example, for a user with 20 interactions, a density of 0.2 results in $20\times0.2=4$ interactions in the training set. According to the experimental results shown in Fig. \ref{inter_density}, we can observe that: (1) As the interaction density increases, the performance of all CDR models shows a rising trend. This is intuitively sensible, as a larger density implies more interaction items, thereby learning more representative user and item embeddings. (2) FUPM consistently outperforms other CDR methods even when the interaction density is very low, which shows that our model effectively alleviate the data-sparsity problem.

\subsubsection{Empirical Study of Privacy (for RQ3)} 
\begin{figure}[!htb]
\centering
\vspace{-5mm}
\subfloat[\scriptsize Book]{\includegraphics[width=1.65in]{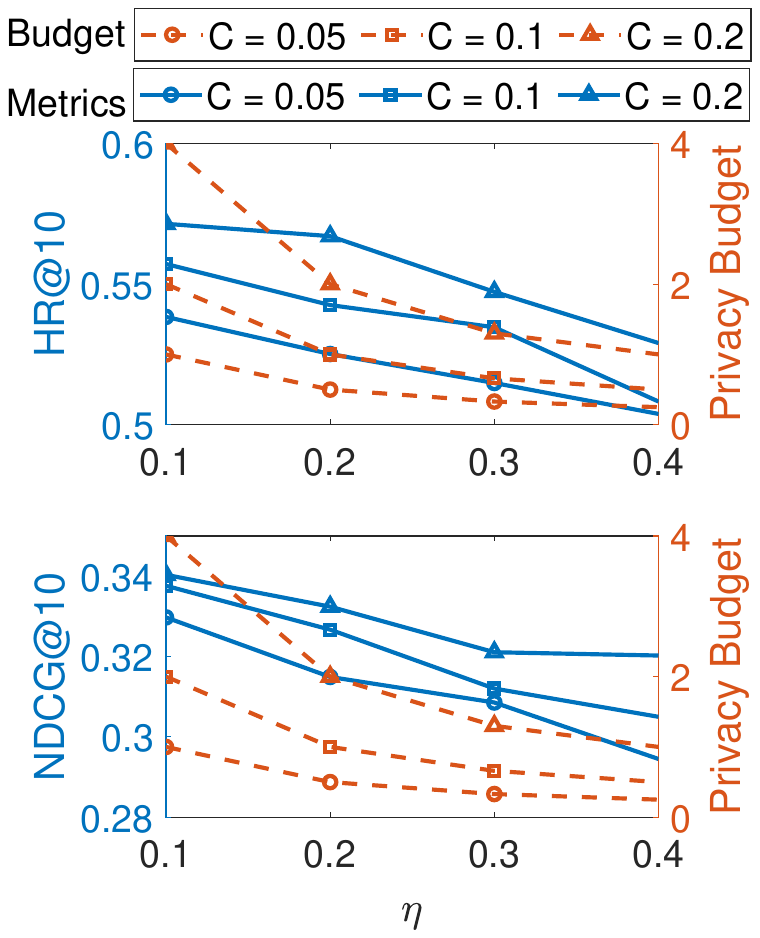}}
\label{book_music_book_privacy}
\hfil
\subfloat[\scriptsize Music]{\includegraphics[width=1.65in]{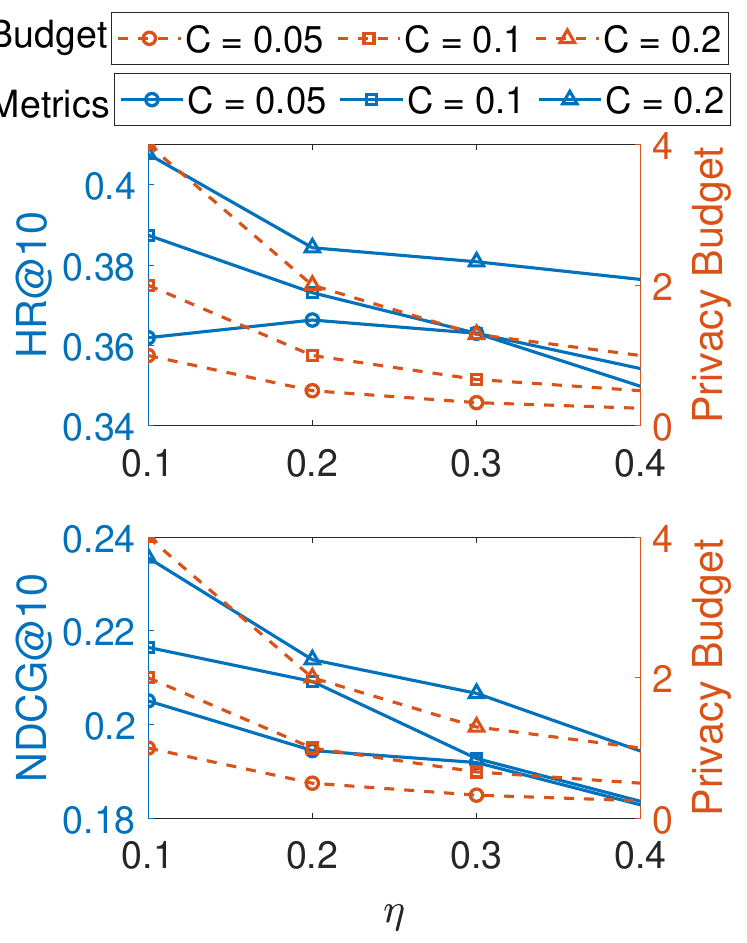}
\label{book_music_music_privacy}}
\hfil
\caption{The performance of different $\eta$ and $C$.}
\label{privacy_analysis}
\vspace{-1mm}
\end{figure}

We compare FUPM and w/o LDP on Task 2 to validate the effectiveness of LDP. 
w/o LDP refers to that we directly transfer prototypes across domains. The results are shown in Table \ref{ldp}. Furthermore, we analyze the trade-off between performance and privacy by varying Laplace noise $\eta$ and clipping threshold $C$, as shown in Fig. \ref{privacy_analysis}. We can find that: (1) Compared with FUPM, w/o LDP achieved better performance, because it leaks more privacy. (2) The noise and clipping threshold jointly determine the model's performance and privacy budget. Larger $\eta$ and smaller $C$ result in a smaller privacy budget but worse performance. Therefore, it's necessary to select moderate values, such as $\eta=0.3$ and $C=0.1$, to balance utility and privacy. 

\begin{table}[htbp]
\centering
\vspace{-3mm}
\caption{The performance impact of LDP.}
\begin{tabular}{ccccc}
%\toprule
\hline
\multirow{2}{*}{Methods} & \multicolumn{2}{c}{Book} & \multicolumn{2}{c}{Music} \\
\cmidrule(lr){2-3}\cmidrule(lr){4-5}
 & HR@10 & NDCG@10 &HR@10&NDCG@10 \\
\midrule
w/o LDP & 0.6044 &0.3835& 0.4900&0.2930 \\
FUPM & 0.5188 &0.3076&0.3630&0.1928  \\ \hline
%\bottomrule
\end{tabular}
\label{ldp}
\end{table}

% \begin{table}[htbp]
% \centering
% \caption{The performance impact of LDP.}
% \label{ldp_effect}
% \begin{tabular}{c c c c c c c c c}
% \hline
% Datasets & Metrics & w/o LDP&FPUPM
%   \\ \hline
% \multirow{3}{*}{Phone} & HR@10 &0.7397&0.7315  \\  
% & NDCG@10 &0.5452&0.5364  \\
% & MRR@10 &0.4846&0.4760  \\ \cline{2-4}
% \multirow{3}{*}{Sport} & HR@10 &0.7283&0.7171  \\  
% & NDCG@10 &0.5384&0.5231  \\
% & MRR@10 &0.4793&0.4629  \\ \hline
% \multirow{3}{*}{Book} & HR@10 &0.6044&0.5347  \\  
% & NDCG@10 &0.3835&0.3120  \\
% & MRR@10 &0.3155&0.2445  \\ \cline{2-4}
% \multirow{3}{*}{Music} & HR@10 &0.4900&0.3696  \\  
% & NDCG@10 &0.2930 &0.1962 \\
% & MRR@10&0.2327&0.1433  \\ \hline
% \end{tabular}
% \end{table}

\vspace{-5mm}
\subsection{Hyper-parameter Analysis (for RQ4)} To optimize recommendation accuracy, we examine the effects of various hyper-parameters on Tasks 1 and 2. Specifically, we investigate the weight parameters $\gamma$ for intra-CL loss and $\alpha$ for inter-CL loss, the aggregation method for global prototypes, and the number of potentially positive items $T$. %and the distributions for generating interpolation coefficients. 

\begin{figure}[!htb]
\centering
\vspace{-5mm}
\subfloat[\scriptsize Phone\&Sport]{\includegraphics[width=1.65in]{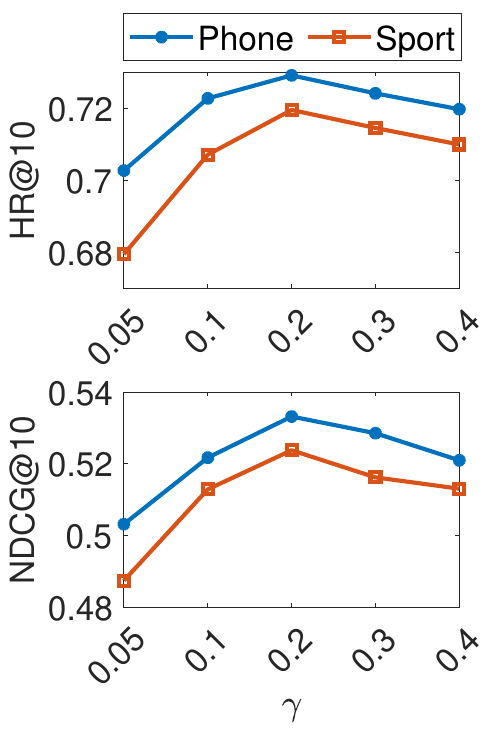}}
\label{phone_sport_gamma}
\hfil
\subfloat[\scriptsize Book\&Music]{\includegraphics[width=1.65in]{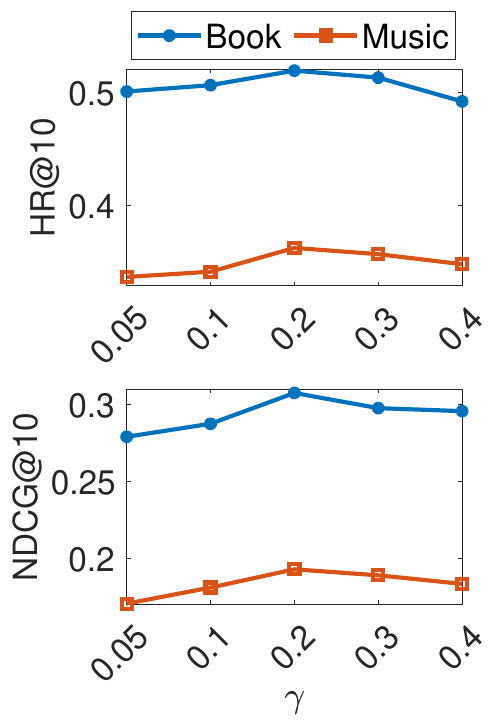}
\label{book_music_gamma}}
\hfil
\caption{The performance of different $\gamma$.}
\label{gamma}
\vspace{-1mm}
\end{figure}

(1) \textbf{Impact of $\gamma$}: The parameter $\gamma$ aims to balance the main loss and the intra-CL loss. It plays a crucial role in modeling comprehensive user preferences. To understand its impact, we vary $\gamma \in \{0.05,0.1,0.2,0.3,0.4\}$ and report the experimental results in Fig. \ref{gamma}. We observe that FUPM achieves the best results when $\gamma$ is set to 0.2. The performance initially rises, peaking at $\gamma=0.2$, and then declines as $\gamma$ increases further. 

\begin{figure}[!htb]
\centering
\vspace{-1mm}
\subfloat[\scriptsize Phone\&Sport]{\includegraphics[width=1.65in]{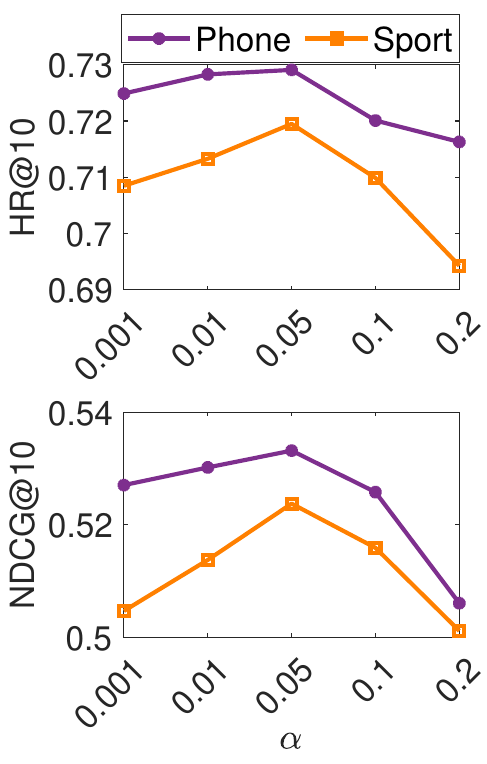}}
\label{phone_sport_alpha}
\hfil
\subfloat[\scriptsize Book\&Music]{\includegraphics[width=1.65in]{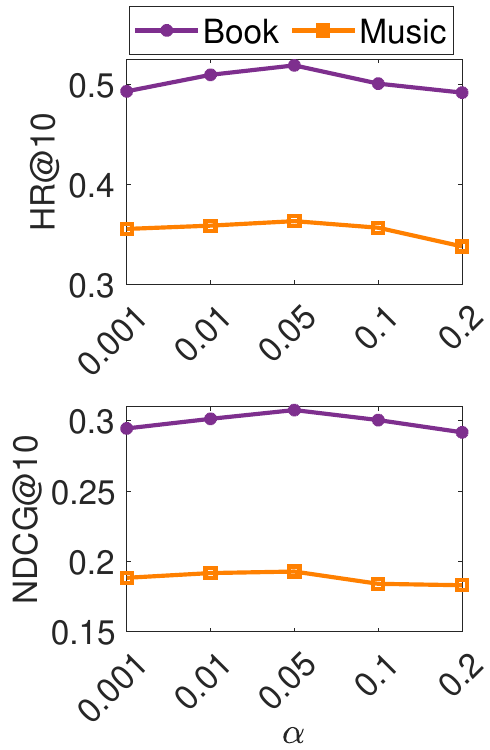}
\label{book_music_alpha}}
\hfil
\caption{The performance of different $\alpha$.}
\label{alpha}
\vspace{-1mm}
\end{figure}

(1) \textbf{Impact of $\alpha$}: The parameter $\alpha$ is crucial for balancing the main loss and the inter-CL loss. It significantly enhances recommendation performance by facilitating effective knowledge transfer across domains through CL. We experiment with $\alpha$ values in the range $[0.001,0.01,0.05,0.1,0.2]$. As shown in Fig. \ref{alpha}, as $\alpha$ increases, the performance first increases and then decreases, reaching the highest value at $\alpha=0.05$.

\begin{figure}[!htb]
\centering
\subfloat[\scriptsize Phone\&Sport]{\includegraphics[width=1.65in]{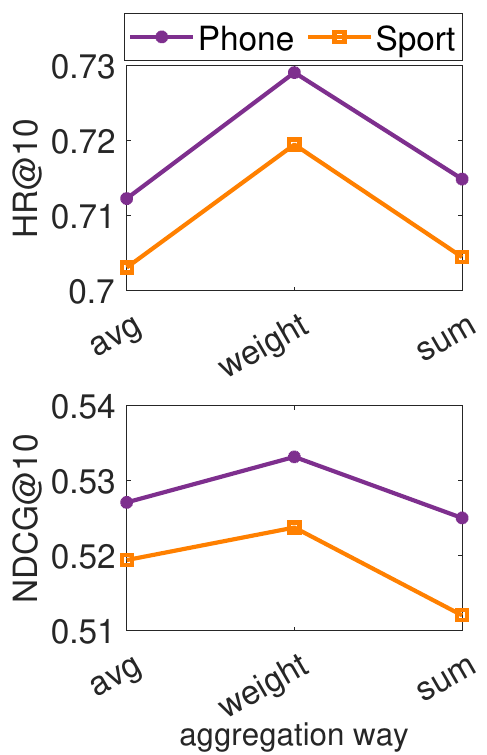}}
\label{phone_sport_proto_agg_way}
\hfil
\subfloat[\scriptsize Book\&Music]{\includegraphics[width=1.65in]{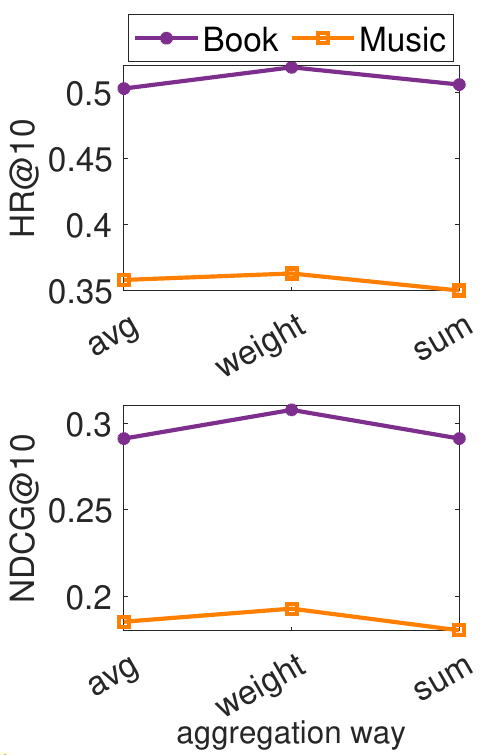}
\label{book_music_proto_agg_way}}
\hfil
\caption{The performance of different prototype aggregation method.}
\label{proto_agg_way}
\end{figure}

(3) \textbf{Impact of aggregation methods for global prototypes}: The aggregation method for global prototypes influences the learning of cross-domain knowledge. To achieve optimal performance, we employ three techniques: element-wise sum, averaging, and weighted fusion. As shown in Fig. \ref{proto_agg_way}, the weight fusion method outperforms the others. This is because weighted fusion adaptively combines local prototypes based on each domain's significance, effectively mitigating the risk of overfitting to sparse domains.

\begin{figure}[!htb]
\centering
\subfloat[\scriptsize Phone\&Sport]{\includegraphics[width=1.65in]{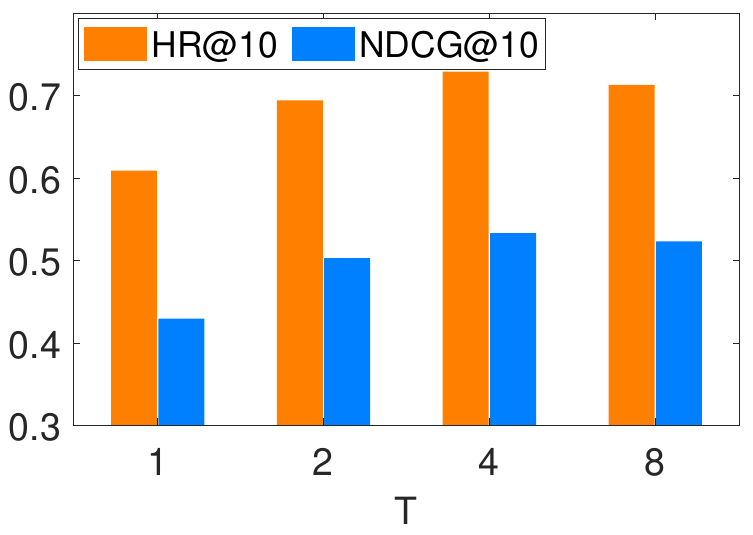}}
\label{phone_sport_phone_potential_item_num}
\hfil
\subfloat[\scriptsize Book\&Music]{\includegraphics[width=1.65in]{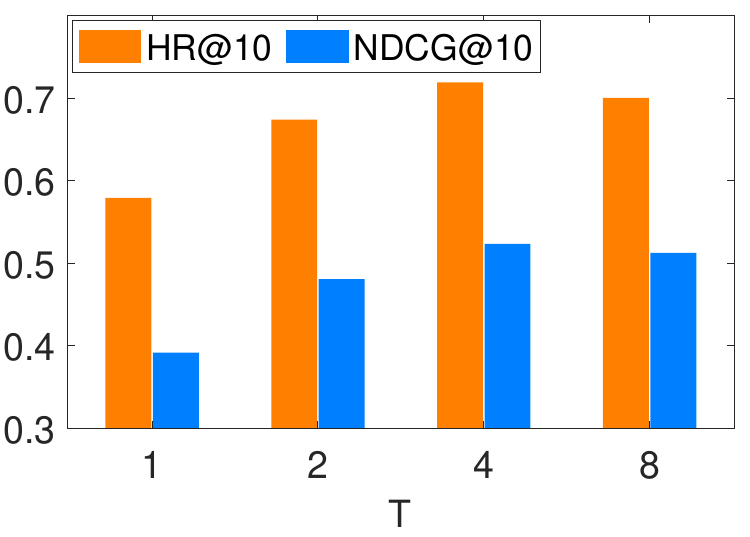}
\label{phone_sport_sport_potential_item_num}}
\hfil
\subfloat[\scriptsize Book\&Music]{\includegraphics[width=1.65in]{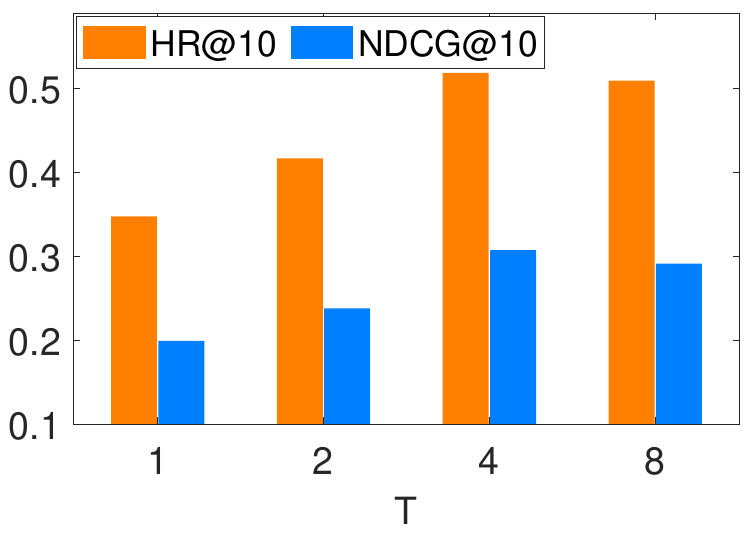}
\label{book_music_book_potential_item_num}}
\hfil
\subfloat[\scriptsize Book\&Music]{\includegraphics[width=1.65in]{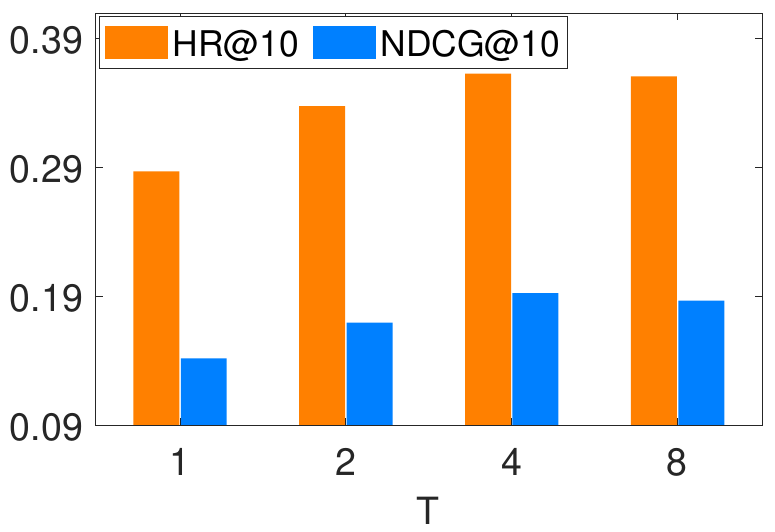}
\label{book_music_music_potential_item_num}}
\hfil
\caption{The performance of different prototype aggregation method.}
\label{potential_pos_item_num}
\end{figure}

(3) \textbf{Impact of the number of potentially positive items}: The number of potentially positive items $T$ plays an important role in the potential interest mining module. To investigate its effect, we experiment with varying numbers of potentially positive items. Based on the results shown in Fig. \ref{potential_pos_item_num}, FUPM achieves the best performance when $T$ is set to 4.

% \begin{figure}[!htb]
% \centering
% \vspace{-5mm}
% \subfloat[\scriptsize Phone\&Sport on Phone]{\includegraphics[width=1.68in]{phone_sport_phone_potential_item_num.pdf}}
% \hfil
% \subfloat[\scriptsize Phone\&Sport on Sport]
% {\includegraphics[width=1.68in]{phone_sport_sport_potential_item_num.pdf}}
% \hfil
% \subfloat[\scriptsize Book\&Music on Book]{\includegraphics[width=1.68in]{book_music_book_potential_item_num.pdf}}
% \hfil
% \subfloat[\scriptsize Book\&Music on Music]
% {\includegraphics[width=1.68in]{book_music_music_potential_item_num.pdf}}
% \hfil
% \caption{The performance of different potential positive item number $T$.}
% \label{potential_pos_item_num}
% \vspace{-1mm}
% \end{figure}

% \begin{figure}[!htb]
% \centering
% \subfloat[\scriptsize Phone\&Sport on Phone]{\includegraphics[width=1.65in]{phone_sport_phone_interpolation.eps}}
% \label{phone_sport_phone_interpolation}
% \hfil
% \subfloat[\scriptsize Phone\&Sport on Sport]
% {\includegraphics[width=1.65in]{phone_sport_sport_interpolation.eps}}
% \label{phone_sport_sport_interpolation}
% \hfil
% \subfloat[\scriptsize Book\&Music on Book]{\includegraphics[width=1.65in]{book_music_book_interpolation.eps}
% \label{book_music_book_interpolation}}
% \hfil
% \subfloat[\scriptsize Book\&Music on Music]
% {\includegraphics[width=1.65in]{book_music_music_interpolation.eps}
% \label{book_music_music_interpolation}}
% \hfil
% \caption{The performance of different distributions.}
% \label{interpolation}
% \end{figure}

\section{Conclusion and Future work}
In this paper, we propose a Federated User Preference Modeling framework (FUPM) for PPCDR to address the data sparsity issue while protecting user privacy. Within FUPM, we first design a comprehensive preference exploration module to learn comprehensive user preferences from review texts and potentially positive items. 
%We employ CL to align ID embeddings with review embeddings, enhancing the representativeness of the learned embeddings. We then utilize potential positive items, identified from uninteracted items that users might be interested in, to explore users' potential interests. 
We then devise a private preference transfer module to privately transfer user preferences within the FL framework. Importantly, in order to protect user privacy during cross-domain knowledge transfer, we learn local prototypes and apply LDP techniques to them before transfer. Extensive experimental results on four CDR tasks based on real-world Amazon and Douban datasets demonstrate the effectiveness of our proposed FUPM. 

Our study assumes overlapping users and non-overlapping items across domains. While our method can be extended to scenarios with partially overlapping users, it is not applicable to scenarios with no user overlap or only partial item overlap. Future work includes exploring effective methods to address these challenges.

\bibliographystyle{IEEEtran}
\bibliography{references}

\vfill

\end{document}